\documentclass{PoS}

\usepackage{amsmath,latexsym,amssymb,slashed}

\usepackage[latin1]{inputenc}
\usepackage[T1]{fontenc}
\usepackage{slashed}

\newcommand{\eq}[1]{\begin{equation}#1\end{equation}}
\newcommand{\spl}[1]{\begin{split}#1\end{split}}

\newcommand{\gcal}{\mathcal{G}}
\newcommand{\kcal}{\mathcal{K}}
\newcommand{\ncal}{\mathcal{N}}
\newcommand{\mcal}{\mathcal{M}}

\newcommand{\gc}{\mathring{g}}
\newcommand{\Oc}{\mathring{\Omega}}

\newcommand{\swed}{{\scriptscriptstyle \wedge}}

\def\bea{\begin{eqnarray}}
\def\eea{\end{eqnarray}}

\def\nn{\nonumber}

\newcommand{\boxedeq}[1]{
\begin{equation}
\fbox{
\rule[0.7cm]{0pt}{0pt}
$#1$
\rule[-0.45cm]{0pt}{0pt}
}
\end{equation}
}

\def\d{\text{d}}

\title{Consistent truncation on Calabi-Yau\\ 
		and Nearly-K\"{a}hler manifolds}

\ShortTitle{Short Title for header}

\author{\speaker{Dimitrios Tsimpis} \\
        {\it   Institut de Physique des Deux Infinis }\\
		{\it Universit\'e de Lyon, UCBL, UMR 5822, CNRS/IN2P3 }\\
		{\it 4 rue Enrico Fermi, 69622 Villeurbanne Cedex, France  }\\
        E-mail: \email{tsimpis@ipnl.in2p3.fr}}

\abstract{We complete and extend the analysis of  arXiv:1903.10504 in several directions: 
we put the 4d theory, arising from the IIA consistent truncation of the universal sector of Calabi-Yau compactification,  
in a form manifestly consistent with 4d $\mathcal{N}=2$ supergravity. 
We go beyond the universal sector and  construct the 4d effective action of IIA compactified on Calabi-Yau's with $h^{1,1}=1$, $h^{2,1}\geq1$,  in the presence of background  flux and fermionic condensates. 
For ALE gravitational instantons, 
we show that the putative quartic gravitino condensate is non-negative, as required for the 
existence of (formal) de Sitter solutions of the 4d theory. 
We discuss some of the issues in promoting these formal solutions to full-fledged string theory de Sitter solutions. 
We also extend the Nearly-K\"{a}hler consistent truncation of 
arXiv:1810.06344 to the complete bosonic sector of one  vector multiplet and one hypermultiplet. }

\FullConference{Corfu Summer Institute 2019 "School and Workshops on Elementary Particle Physics and Gravity" (CORFU2019)\\
		31 August - 25 September 2019\\
		Corfu, Greece}

\begin{document}

\section{Introduction}

 A consistent truncation (CT) of a higher-dimensional theory $S$ is a lower-dimensional theory $S'$ with the property that all solutions  of   $S'$  
  also lift to solutions of $S$.  
 CT's have been constructed on homogeneous spaces \cite{deWit:1986oxb,Nastase:1999cb,Nastase:1999kf,Lu:1999bw,Cvetic:2000nc,Guarino:2015jca}, 
 using a variety of different approaches:  left-invariant forms on  cosets \cite{MuellerHoissen:1987cq,Kapetanakis:1992hf,Bena:2010pr,Cassani:2011fu,Cassani:2012pj},  
 $G$-structures  
\cite{Gauntlett:2007ma,Cassani:2009ck,Cassani:2010uw,Skenderis:2010vz,Gauntlett:2010vu,Liu:2010sa}, 
 double field theory and  exceptional generalized geometry  \cite{Aldazabal:2011nj,Geissbuhler:2011mx,Lee:2014mla,Hohm:2014qga,Cassani:2016ncu,Baguet:2015sma,Baguet:2015iou,Ciceri:2016dmd,Inverso:2016eet,Inverso:2017lrz,Malek:2017njj,Malek:2018zcz}. 
 
 In \cite{Terrisse:2018qjm,Terrisse:2019usq} we initiated the study of CT's in the presence of fermionic condensates. As we review in the following, this approach offers a way to fix the dependence of the condensates on the moduli of the theory. This can be interesting for scenarios where such condensates play an important role, notably in potentially generating a positive cosmological constant.

In \cite{Terrisse:2018qjm} we constructed 	a universal CT on Nearly-K\"{a}hler (NK) manifolds in the presence of dilatino condensates. ``Universal''  means
that the CT  is valid for any NK manifold, independently of whether it is homogeneous, or whether it admits a coset description.\footnote{There exist four homogeneous compact NK six-manifolds: $S^6$, $S^3\times S^3$, $\mathbb{CP}^3$ and $\mathbb{F}^{1,2}$ \cite{2006math.....12655B}, and these were 
until recently the only known compact NK six-manifolds.   In \cite{Foscolo:2015vqa} two non-homogenous examples were constructed, and it 
is has been argued that many more  should exist \cite{2010arXiv1011.4681S}.}  
CT's on homogeneous NK manifolds have been constructed before (in the absence of condensates), relying on the coset description of these spaces \cite{Cassani:2009ck}. Moreover, a universal reduction on NK spaces leading to four-dimensional $\mathcal{N}=2$  gauged supergravity, has been performed in \cite{KashaniPoor:2007tr}, 
but without a proof of consistency.\footnote{
The author of  \cite{KashaniPoor:2007tr} follows the  approach 
of  \cite{Gurrieri:2002wz, DAuria:2004kwe, House:2005yc, Grana:2005ny, Louis:2006kb, KashaniPoor:2006si}, whereby 
one postulates the existence  of a certain finite set of forms on the internal manifold,  and constructs 
a reduction  ansatz based on these forms. Inserting the ansatz 
in the higher-dimensional action 
and integrating over the internal directions 
can then be shown to  result in a lower-dimensional  gauged supergravity. 
In general, this procedure  is not guaranteed to lead to a CT  \cite{KashaniPoor:2006si, Andriot:2018tmb}. 
In addition to showing the compatibility of the reduction with gauged supergravity, \cite{KashaniPoor:2007tr} proves that certain 
supersymmetric solutions of the 4d theory uplift to a well-known class of solutions \cite{Behrndt:2004km,Lust:2004ig} of the 10d theory.
}  In the case where the NK manifold is topologically an $S^6$, the consistency of \cite{KashaniPoor:2007tr} was shown in \cite{Guarino:2015vca}.

In \cite{Terrisse:2019usq} we constructed a  universal CT  of IIA compactified on Calabi-Yau (CY) manifolds, in the presence of background flux and gravitino condensates, comprising the gravity multiplet, one vectormultiplet, and one hypermultiplet. The effective 4d action resulting from type II  compactification on CY manifolds 
has been constructed  before both in the absence \cite{Ferrara:1989ik,Bodner:1990zm} and in the presence of background flux \cite{Louis:2002ny}, and the consistency of the resulting 4d theory with (gauged) $\mathcal{N}=2$ supergravity has been established \cite{Michelson:1996pn, Theis:2003jj,DallAgata:2003sjo,DAuria:2004yjt}. Beyond the universal sector,  these 4d effective actions may not be CT's in general.

In the present paper 
we complete and extend the analysis of  \cite{Terrisse:2018qjm, Terrisse:2019usq} 
in several directions. In section \ref{sec:universal} 
we put the 4d theory, arising from the IIA consistent truncation of the universal sector of Calabi-Yau compactification,  
in a form manifestly consistent with 4d $\mathcal{N}=2$ supergravity. In section \ref{sec:cycomp}  
we go beyond the universal sector and  construct the 4d effective action of IIA compactified on Calabi-Yau's with $h^{1,1}=1$, $h^{2,1}\geq1$,  in the presence of background  flux and fermionic condensates. 
The 4d actions thus derived are valid for generic gravitino condensates arising from gravitational instanton backgrounds with positive-chirality 4d zero modes. 
Specializing to the case of ALE gravitational instantons, in section \ref{sec:positivity} 
we show that the putative quartic gravitino condensate is non-negative, as required for the 
existence of (formal) de Sitter solutions of the 4d theory \cite{Terrisse:2019usq}.  
We discuss some of the issues in promoting these formal solutions to full-fledged string theory de Sitter solutions in section \ref{sec:conclusions}. 
In section \ref{sec:nk} we also extend the Nearly-K\"{a}hler consistent truncation of 
\cite{Terrisse:2018qjm} to the complete bosonic sector of one  vector multiplet and one hypermultiplet.

\section{CY universal truncation}\label{sec:universal}

Let us recall the consistent truncation of \cite{Terrisse:2019usq}: it is obtained from massless IIA supergravity  
by the following form ansatz, 
cf.~appendices \ref{app:spin}, \ref{sec:the10d} for our conventions, 
\eq{\spl{\label{foranscyb}
F&=\d\alpha~;~~~ H=\d\chi \swed J+\d\beta+\frac12\text{Re}\big(b_0\Omega^*\big)\\
G&=\varphi\text{vol}_4+\frac12 c_0J\swed J+ J\swed (\d\gamma - \alpha\wedge \d\chi)-\frac{1}{2}D\xi\swed\text{Im}\Omega
-\frac{1}{2}D\xi'\swed\text{Re}\Omega
~,}}
where  $\alpha$, $\gamma$ are 4d one-forms; $\beta$ is a 4d two-form dual to a scalar $b$ (see \eqref{prhd2qri} below for the precise relation); $\varphi$, $\chi$, $\xi$, $\xi'$ are 4d scalars (of these $\varphi$ turns out to be  auxiliary and will be eliminated by its equations of motion); 
$c_0\in\mathbb{R}$ and $b_0\in\mathbb{C}$ are background fluxes; $J$ and $\Omega$ are the CY K\"{a}hler form and holomorphic three-form respectively. 
The covariant derivatives are defined as follows,
\eq{\label{43i}
D\xi:=\d\xi+b_1\alpha~;~~~D\xi':=\d\xi'+b_2\alpha
~,}
where we set $b_0=ib_1+b_2$. 
The expressions for the forms $F$, $H$, $G$ above are such that the 10d Bianchi identities \eqref{bi} are automatically satisfied for vanishing mass. 
The ansatz for the metric reads,
\eq{\label{tdma}\d s^2_{(10)} =e^{2A(x)}\left(e^{-8A(x)} g_{\mu\nu}\d x^{\mu}\d x^{\nu}+g_{mn}\d y^m\d y^n 
\right)~,
}
where $A$ is a scalar depending only on the 4d coordinates $x^\mu$.  
The 4d field content corresponds to the bosonic sector of $\mathcal{N}=2$, 4d supergravity with 
a gravity multiplet $(g_{\mu\nu}, \alpha)$, one vector multiplet $(\gamma, A, \chi)$, and one hypermultiplet $(\phi,b,\xi,\xi')$.\footnote{
More precisely, the  $\mathcal{N}=2$ multiplets should be expressed in terms of volume modulus $v$, the 4d dilaton $\hat{\phi}$ and the vector-multiplet 
vector $\tilde{\gamma}$,  cf.~section \ref{sec:cmp}.
}

The ansatz is supplemented by 4d gravitino condensates arising potentially from  gravitational instantons supporting spin-3/2 zero modes. 
We start from the following 10d gravitino truncation,
\eq{\label{ans:gravitino}
\Psi_m=0~;~~~\Psi_{\mu+}=\psi_{\mu+}\otimes\eta~; ~~~\Psi_{\mu-}=\psi_{\mu+}'\otimes\eta^c 
~,}
where we have Wick-rotated to Euclidean signature; 
$\eta$ is the (normalized) covariantly-constant spinor associated with the 
CY metric $g_{mn}$; the 4d gravitini $\psi_{\mu+}$, $\psi_{\mu+}'$ are assumed to have positive chirality, as e.g.~in the case of ALE gravitational instantons. They are not dynamical: they should be thought of as 
zero modes of the Dirac operator in the background of the gravitational instanton. 
The condensates are then expressed in terms of two ``parameters'' $\mathcal{A}$,  $\mathcal{B}$ corresponding to the quadratic and quartic condensates respectively,\footnote{\label{ft1}Note however that $\mathcal{A}$,  $\mathcal{B}$ have a  4d metric dependence, due to the contractions between vector indices of the gravitini.}
\eq{\spl{\label{calbdef}
\mathcal{A}&:= (\tilde{\psi}_{\mu+}\gamma^{\mu\nu}\psi'_{\nu+})
=- (\tilde{\psi}^{\mu}_+\psi'_{\mu+})\\
\mathcal{B}&:= 
 -\frac{3}{2} (\tilde{\psi}_{[\mu}\psi'_{\nu]})^2
 + (\tilde{\psi}^\mu \gamma^{\rho\nu} \psi'_\mu) (\tilde{\psi}_\rho \psi'_\nu) 
 +3 (\tilde{\psi}_{[\mu_1} \gamma_{\mu_2\mu_3}\psi'_{\mu_4]})^2
~.}}
We emphasize that these expressions are valid for all  gravitational instantons with positive-chirality gravitino zero modes: they are {\it not} limited to ALE gravitational instantons.

As was shown in \cite{Terrisse:2019usq}, substituting the truncation ansatz \eqref{foranscyb}-\eqref{calbdef} into  the 10d equations of motion of appendix \ref{sec:the10d}, results in a set of 
4d equations, all of which are derivable from a single 4d action, thereby  proving the consistency of the truncation. 
The  4d action is given by, 
\boxedeq{\spl{\label{ctr2}
S_4=\int&\d^4 x\sqrt{g}
\Big(
R
 - 24 (\partial A)^2 
 -\tfrac{1}{2} (\partial \phi)^2  
-\tfrac{3}{2}  e^{-4A - \phi}(\partial \chi)^2 
- \tfrac{1}{2} e^{-6A + \phi/2} \left[(D \xi)^2+(D \xi')^2\right]\\
&-\tfrac{1}{2} e^{\phi - 12A} (D b )^2
-\tfrac{1}{4} e^{3\phi/2 + 6A} \d\alpha^2
-\tfrac{3}{4}  e^{\phi/2 + 2A} (\d\gamma - \alpha\swed\d\chi)^2
-V
 \Big) 
%CS terms:
+\int 
3 \chi\ \d\gamma\swed\d\gamma  
~,
}}
where we have dualized the two-form $\beta$ to a scalar $b$, the ``axion'', via the relation, 
\eq{\label{prhd2qri}
 \d \beta =e^{\phi-12A}\star_4 D b~;~~~
 D b:=
 \left[\d b+\tfrac12(\xi \d\xi'-\xi'\d\xi) + 3c_0(\gamma - \chi \alpha)+\Xi\alpha
\right]
~,}
with $\Xi:=b_2\xi-b_1\xi'$. The potential of the theory is given by,
\boxedeq{\spl{\label{58}
V(\chi,\xi,\xi',\phi,A)
=
&\tfrac{3}{2}  c_0^2e^{\phi/2 - 14A}
 + \tfrac{1}{2}|b_0|^2e^{-\phi-12A}
\\
-&3c_0 {\mathcal{A}}e^{\phi/4 - 4A}
+e^{6A}\mathcal{B} +\tfrac12\Big(
{\mathcal{A}}e^{3A}+(3c_0\chi -\Xi)e^{-\phi/4 - 9A}
\Big)^2
~.
}}
Note that   for $c_0\neq0$, \eqref{ctr2}-\eqref{58} are invariant under the 
shift, 
\eq{\label{shiftui}
\chi\rightarrow\chi+\frac{e_0}{3c_0}~;~~~\Xi\rightarrow\Xi+e_0
~,}
where $e_0$ is an arbitrary real constant. Interestingly, setting the background flux $b_0, c_0$ to zero  {\it after} performing the shift \eqref{shiftui}, 
still results in a gauged theory provided $e_0\neq 0$ \cite{Louis:2002ny}.

{\it Euclidean  signature}

Instantons are solutions of the action in Euclidean signature. 
However, as is well-known,  the operation of dualization does not commute with Wick rotation \cite{Giddings:1987cg}. 
Starting from the Euclidean action 
in terms of the two-from $\beta$  and {\it then} dualizing to the axion $b$, one obtains 
an action which differs form \eqref{ctr2} in the sign of the kinetic term for $b$.

\subsection{Comparison with $\mathcal{N}=2$, 4d supergravity}\label{sec:cmp}

Let us introduce a complex scalar $t$ and the real scalars $v$, $\hat{\phi}$ defined as follows,
\eq{\label{eqti}
t:=\chi+i v~;~~~
v:=e^{2A+\phi/2}
~;~~~\hat{\phi}:=
\tfrac{1}{4}\phi-3A
~.}
As we will see in the following  $v$ and $\hat{\phi}$ will be identified with the volume modulus and the 
4d dilaton respectively. 
Moreover we introduce a new vector $\tilde{\gamma}$ and its field-strength,
\eq{\label{svecs}
\tilde{\gamma}:= \gamma-\chi\alpha~;~~~
\tilde{\mathcal{F}}:=\d\tilde{\gamma}
~,}
so that $\d\gamma-\alpha\swed\d\chi=\tilde{\mathcal{F}}+\chi \mathcal{F}$, 
where $\mathcal{F}:=\d\alpha$. 
In terms of the new fields, 
the action    \eqref{ctr2}  reads,
\boxedeq{\spl{\label{actionfinal2i}
S_4=\int&\d^4 x\sqrt{g_4}
\Big(
R
 - 2\partial_t\partial_{\bar{t}}K_V|\d t|^2 
-2(\d\hat{\phi})^2- \tfrac{1}{2} e^{4\hat{\phi}} (D b)^2
- \tfrac{1}{2} e^{2\hat{\phi}} \left| D\xi'+i D\xi\right|^2
\\
&-\tfrac{1}{4} v^{3} \mathcal{F}^2
-\tfrac{3}{4}  v (\tilde{\mathcal{F}}+\chi \mathcal{F})^2
-V
 \Big) 
+\int 
3 \chi \tilde{\mathcal{F}}\swed\tilde{\mathcal{F}}
+ \chi^3  {\mathcal{F}}\swed {\mathcal{F}}
+3\chi^2  {\mathcal{F}}\swed\tilde{\mathcal{F}}
~,
}}
where we have introduced the vector-multiplet K\"{a}hler potential,
\eq{\label{kpot1i}
K_V=-3\ln(t-\bar{t})+\text{const}
~.
}
It can be verified that 
the action \eqref{actionfinal2i} is of the form of $\mathcal{N}=2$ supergravity, cf.~\eqref{n2}. 
To see this, first note that there is only one complexified K\"{a}hler modulus, so  that  $I=0,1$ in the formulae of Appendix \ref{sec:gaugedsugra}. Accordingly we may drop the $i$-index in  \eqref{99}, so that the 
$t^i$ variable therein is identified with the $t$ of  \eqref{kpot1i}. 
We may choose the unique  harmonic (1,1)-form on the CY  so that the triple intersection is normalized to  $\kcal_{ijk}=6$, cf.~\eqref{100}.  
Taking \eqref{eqti} into account, we then find $\kcal=v^{3}$, so that  \eqref{101} precisely reduces to \eqref{kpot1i}, and the vector-multiplet scalar kinetic terms of 
\eqref{n2} reduce to those of  \eqref{actionfinal2i}. 
Moreover, taking \eqref{eqti} into account, eq.~\eqref{37} gives,
\eq{
\Re\ncal= \left(
\begin{matrix}
    2\chi^3      & 3\chi^2 \\
    3\chi^2  & 6\chi
\end{matrix}\right)
~;~~~
\Im\ncal=
\left(
\begin{matrix}
   v^{3}+3v \chi^2      & 3v \chi  \\
   3v \chi   &  3v
\end{matrix}\right)
~.}
Substituting in eq.~\eqref{n2} then reproduces precisely the gauge kinetic and Chern-Simons terms in \eqref{actionfinal2i}, provided we identify $\alpha$, $\tilde{\gamma}$ with $A^{I=0}$, $A^{I=1}$ respectively. The corresponding Killing vectors can then be read off by comparing \eqref{43i}, \eqref{prhd2qri} with \eqref{43prime},
\eq{\label{kilaci}
k^u_{I=0}\partial_{q^u}=-\Xi\partial_{b}-b_1\partial_{\xi}-b_2\partial_{\xi'}~;~~~
k^u_{I=1}\partial_{q^u}=-3c_0\partial_{b}
~,}
where the  $q^u$ coordinates are identified with $\hat{\phi}$, $b$,  $\xi$, $\xi'$, and we took into account that there is only one hypermultiplet: $n_H=1$,  
so that $u=1,\dots, 4$. 
To compare with the quaternionic metric \eqref{qmetric} first note that there are no complex structure deformations, so that the $z^a$'s are absent in our case, 
and the hypermultiplet indices take on  a single value, $A,B=0$. The periods are then calculated from $\gcal=-\tfrac{i}{2}(Z^0)^2$, evaluated at $Z^0=1$, 
so that $\Omega=\alpha+i\beta$ in terms of the basis in \eqref{3basis}. This then gives 
$\mcal_{AB}=-i$, and \eqref{qmetric} precisely reduces to the kinetic $\hat{\phi}$, $b$, $\xi$, $\xi'$ terms in \eqref{actionfinal2i}. Finally, to show that the potential \eqref{58} is of the from 
given in \eqref{potac}, we compute,
\eq{\spl{\label{32i}
h_{uv}k_I^uk_J^v=\frac12 \left(
\begin{matrix}
   e^{4\hat{\phi}}\Xi^2+ e^{2\hat{\phi}}|b_0|^2)   & -3c_0e^{4\hat{\phi}}\Xi \\
   -3c_0e^{4\hat{\phi}}\Xi   & 9c_0^2e^{4\hat{\phi}}
\end{matrix}\right)
~,}}
and,
\eq{\label{32bi}
X^{(I}\bar{X}^{J)}
=\left(
\begin{matrix}
  1    &  \chi\\
   \chi  & \chi^2+ v^{2}
\end{matrix}\right)
~;~~~
\kcal(\Im \ncal^{-1})^{IJ}
=-\left(
\begin{matrix}
  1    &  \chi\\
   \chi  & \chi^2+\frac13 v^{2}
\end{matrix}\right)
~.}
Moreover, taking \eqref{prep1} into account,  \eqref{kilaci}, \eqref{vecp} imply,
\eq{\label{33i}
\vec{P}_{I=1}=(0,0,-\frac{3}{\sqrt{2}}c_0e^{2\hat{\phi}})
~.}
Substituting \eqref{32i}-\eqref{33i} into \eqref{potac} reproduces precisely the potential given in \eqref{58}, in the case of vanishing condensates: $\mathcal{A}=\mathcal{B}=0$.

\section{CY compactification with $h^{1,1}=1$, $h^{2,1}\geq1$}\label{sec:cycomp}

We will now extend the universal sector truncation of section \ref{sec:universal}, to include an arbitrary number of hypermultiplets. 
In order to obtain the 4d theory we will simply insert the new ansatz, eqs.~\eqref{tdmai}, \eqref{ans:metric}, \eqref{ans:forms} below, into the 10d action, keeping up to and including quadratic terms in the fluctuations. 
This is the standard procedure  in the literature when it comes to CY compactification: it should lead to a low-energy effective action, 
but not necessarily to a consistent truncation \cite{KashaniPoor:2006si, Andriot:2018tmb}. 
Nevertheless we will see that it does satisfy certain consistency requirements: (a) the resulting 4d Lagrangian contains the consistent truncation of   \cite{Terrisse:2019usq} as a subsector, plus terms that depend on the additional hypermultiplet scalars of the present paper; (b) the kinetic terms of the additional  hypermultiplet scalars are of the standard form; (c) the resulting Lagrangian can be cast in the form of the bosonic sector of a theory with manifest $\mathcal{N}=2$, $d=4$ supergravity.

The ansatz for the ten-dimensional metric reads,
\eq{\label{tdmai}\d s^2_{(10)} =e^{2A(x)}\left(e^{-8A(x)} g_{\mu\nu}(x)\d x^{\mu}\d x^{\nu}+g_{mn}(x,y)\d y^m\d y^n 
\right)~,
}
where the scalar $A$ only depends on the four-dimensional coordinates $x^\mu$, but not on the $y^m$ coordinates of the CY. 
Contrary to  the consistent truncation ansatz of \cite{Terrisse:2019usq}, the internal CY metric $g_{mn}$ is now allowed to vary as we move along the four-dimensional spacetime. Explicitly we set,
\eq{\label{ans:metric}
g_{mn}(x,y):=\gc_{mn}(y)+\delta g_{mn}(x,y)~;~~~ \delta g_{mn}(x,y):=\frac{6}{|\Oc|^2}\left(
\delta z^a(x) (\Phi_a)_{m}{}^{pq}\Oc^*_{npq}+\text{c.c.}
\right)~,
}
where $a=1,\dots,h^{2,1}$. 
The metric is thus expressed as a fluctuation around a  fiducial CY metric $\gc_{mn}(y)$. 
We may think of the latter as being defined at a 
point $\mathring{z}^a$ in the  (complex) $h^{2,1}$-dimensional space of complex structure moduli $\mathfrak{M}_c$ , while 
the nearby metric ${g}_{mn}$ is defined at the  
point  $\mathring{z}^a+\delta z^a$. 
The variations $\delta z^a$ span the cotangent space of $\mathfrak{M}_c$ at 
the point $\mathring{z}^a$, and 
the fluctuation $\delta g_{mn}(x,y)$ is a complex-structure deformation. 
The $\Phi_a$'s, $a=1,\dots, h^{2,1}$, consitute a basis of  harmonic (2,1)-forms, with respect to the metric and complex structure at the point $\mathring{z}^a$.\footnote{\label{f2}The metric deformation can be viewed as resulting from a variation of the holomorphic top form, 
\eq{\delta\Omega=\delta z^a \Phi_a~,\nn}
with the induced deformation of complex structure given by,
\eq{
\delta\mathcal{I}_m{}^n=\frac{6i}{|\Omega|^2}\left(
\delta z^a(\Phi_a)_{mpq}\bar{\Omega}^{ npq}
-\delta \bar{z}^{a}(\bar{\Phi}^a)^{ npq}\Omega_{mpq}
\right)
~.\nn}
The associated metric variation, $\delta g_{mn}=-\delta \mathcal{I}_m{}^q J_{qn}$, is then given by \eqref{ans:metric}. Note that 
the right-hand side of that equation is automatically symmetric in $(m,n)$.}  
Furthermore we will assume that the volume of the metric $g_{mn}$ is constant, so that the volume modulus of the compactification space is   given by 
the scalar $v$ of \eqref{eqti}, as in the previous section. 
The complete moduli space $\mathfrak{M}$ of the compactification has a 
direct-product structure: 
\eq{\label{fia}
\mathfrak{M}=\mathfrak{M}_k\times \mathfrak{M}_c~,} 
where $\mathfrak{M}_k$ is  the (real) one-dimensional moduli space of K\"{a}hler deformations  
parametrized  by the volume modulus.

We denote by $(\mathring{J},\mathring{\Omega})$ the K\"{a}hler and holomorphic forms of the fiducial CY, defined at 
the point  $\mathring{z}^a$, while those of the nearby CY, defined at 
the point  $\mathring{z}^a+\delta z^a$, will be denoted by $(J,\Omega)$.\footnote{\label{f1}In the present paper, to conform with 
the standard treatment of complex structure deformations, 
 we do not impose a normalization on $\Omega$. 
Compared to the 
conventions of \cite{Terrisse:2019usq}: $\Omega^{\text{there}}=\frac{4\sqrt{3}}{|\Omega^{\text{here}}|}\Omega^{\text{here}}$.} 
Since the CY metric $g_{mn}$  is assumed to have fixed volume, and its associated K\"{a}hler form does not change under complex structure deformations, we have $\mathring{J}=J$. 
It can be seen that $\delta g_{mn}$ given in \eqref{ans:metric} is a solution to the six-dimensional Ricci-flatness condition ${R}_{mn}(\gc+\delta g)=0$, to linear order in 
$\delta z^a$.

Plugging the metric ansatz into the 10d Einstein term we obtain,
\eq{\spl{
\label{rs} \sqrt{g_{10}}R_{10}=\sqrt{g_6}\sqrt{g_4}\Big( 
{R}_4&-24(\partial A)^2
-g^{\mu\nu}\hat{\nabla}_{\mu}\big(g^{mn}\partial_{\nu}\delta g_{mn} \big)
\\
&
-\frac14 g^{\mu\nu}\left[ g^{mn}g^{pq}\partial_{\mu}\delta g_{mp}\partial_{\nu}\delta g_{nq}
+ g^{mn}g^{pq}\partial_{\mu}\delta g_{mn}
\partial_{\nu}\delta g_{pq}\right]
\Big)
~,}}
where we have taken into account  the Ricci-flatness of the six-dimensional metric $g_{mn}(x,y)$; $\hat{\nabla}$ is the connection associated to the 4d metric $\hat{g}_{\mu\nu}:=e^{-8A(x)} g_{\mu\nu}$. 
So far equation \eqref{rs} is exact, in that the deformations $\delta g$  do not need to be infinitesimal. 
Moreover the internal metric depends on the 4d spacetime only through the fluctuation, $\partial_{\nu}\delta g_{mn}=\partial_{\nu} g_{mn}$, so that $g^{mn}\partial_{\nu}\delta g_{mn}=\partial_{\nu}\!\ln g_6=0$, where in the last equality we took into account that 
 $g_{mn}$ has fixed volume, cf.~the discussion below \eqref{ans:metric}. 
Since eq.~\eqref{rs} is already quadratic in the fluctuations, we may use the linearized expressions \eqref{ans:metric} for $\delta g_{mn}$. 
To quadratic order in $\delta z^a$, 
eq.~\eqref{rs} then simplifies to,
\eq{
\label{simple}\int\sqrt{g_{10}}R_{10}={V}_6\int\sqrt{g_4}\Big( 
{R}_4-24(\partial A)^2
-2g_{a\bar{b}}\d z^a\d\bar{z}^b
\Big)
~,}
where we have defined,
\eq{\label{csm}
g_{a\bar{b}}:=-\frac{\int\Phi_a\swed\bar{\Phi}_b}{\int\Omega\swed\Omega^*} 
~,}
and we have used that $\partial_\mu\delta z^a=\partial_\mu z^a$. Moreover 
 we have taken \eqref{volstuff2} into account, and introduced the CY volume ${V}_6:=\int\text{vol}_6$ which is constant, 
cf.~the discussion below \eqref{ans:metric}.

Note that as long as we are only keeping quadratic terms in the fluctuations, 
it does not matter whether we use $\Oc$ or  $\Omega$  in  \eqref{csm}. However, using $\Omega$  leads to a covariant expression 
with respect to the moduli. 
Indeed, in accordance with the standard form for the kinetic terms of complex-structure moduli, $g_{a\bar{b}}$ can be viewed as a metric derived from a K\"{a}hler potential,\footnote{Recall that as we move in  $\mathfrak{M}_c$, $\Omega$ varies holomorphically with respect to $z^a$, so that $\partial_{z^a}\bar{\Omega}=0$. Moreover rescalings of the form $\Omega\rightarrow f(z)\Omega$, where  $f$ is an arbitrary holomorphic function of $z^a$,  do not change the complex structure of the CY. Thus $\Omega$ may be viewed as a section of a holomorphic line bundle 
over  $\mathfrak{M}_c$ \cite{Strominger:1990pd}. Motion in  $\mathfrak{M}_c$ 
is described in terms of the K\"{a}hler-covariant derivative:
\eq{
\mathcal{D}_a \Omega=\Phi_a ~;~~~
\delta\Omega=\delta z^a \mathcal{D}_a \Omega
~,\nn}
where $\mathcal{D}_a := \partial_{z^a}+\partial_{z^a} {K}_H$. 
This is consistent with the definition of the K\"{a}hler potential in (\ref{kpot}) as can be seen by wedging both sides of the 
first equation above with $\bar{\Omega}$ and integrating over the CY. The consistency with \eqref{csm} follows similarly.}
\eq{\label{kpot}
g_{a\bar{b}}=\partial_{ z^{a\phantom{b}} }\!\!\bar{\partial}_{ {z}^b} {K}_H
~;~~~
 {K}_H(z,\bar{z})=-\ln\Big(i\int
\Omega\swed\Omega^*
\Big)
~.}
%
%{\it The form terms}
%\label{sec:lagr}
%
Let us now consider the ansatz for the forms. These 
 are expanded so that the 10d Bianchi identities, eq.~\eqref{bi} with zero mass, are automatically satisfied, 
\eq{\spl{\label{ans:forms}
F&=\d\alpha~;~~~ H=\d\chi \swed J+\d\beta+  p^A\alpha_A+q_A\beta^A \\
G&=\varphi\text{vol}_4+\frac12 c_0J\swed J+ J\swed (\d\gamma - \alpha\wedge \d\chi)- \big( D\xi'_A\swed \beta^A
+ D\xi^A\swed\alpha_A\big)
~,}}
where we have 
expanded on the basis \eqref{3basis}, and 
introduced  background charges  $p^A$, $q_A\in\mathbb{R}$, following the notation of \cite{Louis:2002ny}; the index $A$ is related to the index $a$ in \eqref{fia} 
via $A=(0,a)$, so that  $A=0,1,\dots,h^{2,1}$. 
The covariant derivatives are given by, 
\eq{\label{covxi}
D\xi^A:=\d\xi^A+p^A\alpha~;~~~D\xi'_A:=\d\xi'_A+q_A\alpha 
~.}
The following expressions are useful,
\eq{\spl{\label{hodsr}
\star_{10} F &= \frac{1}{6}e^{6A} \star_4\d\alpha\swed {J}^3 
\\	
\star_{10} H &=  \tfrac12 e^{4A+2B} \star_{4}\!\d\chi\swed   {J}^2
+ \tfrac16 e^{4A-2B} \star_{4}\!h\swed   {J}^3+e^{-12A}( p^A\star_{6}\!\alpha_A+ q_A\star_{6}\!\beta^A)\swed\text{vol}_4
\\
\star_{10} G &= -\tfrac16 \varphi e^{2A-4B}  {J}^3
+c_0 e^{2A+4B} \text{vol}_4\swed {J}+\tfrac12 e^{2A}\star_{4}(\d\gamma - \alpha\wedge \d\chi)\swed {J}^2\\
&~~~~\!-\tfrac{1}{2} e^{2A+2B}   \star_{4}\! D\xi'_A\swed  \star_{6}\!\beta^A
-\tfrac{1}{2} e^{2A+2B}   \star_{4}\! D\xi^A\swed  \star_{6}\!\alpha_A
~,}}
where the four-dimensional Hodge-star is with respect to the unwarped metric. 
Moreover, taking \eqref{abstar} into account,  we compute, 
\eq{\spl{
F^2 &= e^{-4A-4B} \d\alpha^2\\
H^2 &= 18e^{-6A-2B}(\partial\chi)^2+e^{-6A-6B}h^2  -6e^{-6A}(q+\mcal\cdot p)\cdot\Im\mcal^{-1}\cdot(q+{\mcal}^*\cdot p)\\
G^2 &= -24 e^{-8A-8B}\varphi^2
-24e^{-8A-2B}(D\xi'+\mcal \cdot D\xi)\cdot\Im\mcal^{-1}\cdot(D\xi'+{\mcal}^*\cdot D\xi)\\
&~~~~\!+72c_0^2e^{-8A}+36e^{-8A-4B}(\d\gamma - \alpha\wedge \d\chi)^2
~,}}
where the contractions on the left-hand sides above are computed with respect to the ten-dimensional metric while the contractions 
on the right-hand sides are taken with respect to the unwarped metric.

Plugging the above expressions into the ten-dimensional action and integrating over the internal directions, we obtain the four-dimensional Lagrangian,
\eq{\spl{\label{blipuy}
S_4=\int\d^4 x\sqrt{g}&
\Big(
R
 - 24 (\partial A)^2 
 -\tfrac{1}{2} (\partial \phi)^2  -2g_{a\bar{b}}\d z^a\d\bar{z}^b
-\tfrac{3}{2}  e^{-4A - \phi}(\partial \chi)^2\\
&+ \tfrac{1}{2} e^{-6A + \phi/2} (D\xi'+\mcal \cdot D\xi)\cdot\Im\mcal^{-1}\cdot(D\xi'+{\mcal}^*\cdot D\xi)\\
&
-\tfrac{1}{4} e^{3\phi/2 + 6A} \d\alpha^2
-\tfrac{3}{4}  e^{\phi/2 + 2A} (\d\gamma - \alpha\swed\d\chi)^2
-\tfrac{1}{12} e^{-\phi + 12A} \d\beta^2-V
 \Big)\\ 
%CS terms:
+\int 
&3c_0 \d(\gamma-\alpha\chi)\swed\beta
+3 \chi\ \d\gamma\swed\d\gamma 
+\Xi\beta\swed\d\alpha 
-\beta\swed D\xi'_A\swed D\xi^A
~,
}}
where the potential of the theory is given by,
\eq{\spl{\label{potentialfinal1}
V(\chi,\phi,A)
&= ( \Xi-3 c_0\chi)\varphi -e^{12A +\phi/4}\varphi\mathcal{A}-\tfrac{1}{2} e^{18A +\phi/2}\varphi^2\\
& 
 +\tfrac{3}{2}  c_0^2e^{\phi/2 - 14A}
 - \tfrac{1}{2}(q+\mcal\cdot p)\cdot\Im\mcal^{-1}\cdot(q+{\mcal}^*\cdot p)e^{-\phi-12A}\\
& 
-3c_0 {\mathcal{A}}e^{\phi/4 - 4A}
+e^{6A} \mathcal{B} 
~,
}}
with  $\Xi:=p^A\xi'_A-q_A\xi^A$; the gravitino condensates were given in \eqref{calbdef}.  

The non-dynamical scalar $\varphi$ 
is constrained by the ten-dimensional $G$-field equation of motion, cf.~\eqref{beomf3}, which implies in particular,  
\eq{\label{gfeom2modqr}
0=\d\left(
 e^{\phi/2+18A} \varphi+e^{\phi/4+12A}\mathcal{A}+3c_0\chi-\Xi
\right)
~.}
This is consistent with the equations of motion for $\varphi$ coming from \eqref{blipuy}, \eqref{potentialfinal1},
\eq{\label{phisol}
\varphi=e^{-\phi/2-18A} \left(\Xi -3c_0\chi-e^{\phi/4+12A}\mathcal{A}\right)
~.}
Plugging the above back into \eqref{potentialfinal1}, we obtain the following 
expression for  the potential,
\boxedeq{\spl{\label{58x}
V(\chi,\xi,\xi',\phi,A)
=
&\tfrac{3}{2}  c_0^2e^{\phi/2 - 14A}
 - \tfrac{1}{2}(q+\mcal\cdot p)\cdot\Im\mcal^{-1}\cdot(q+{\mcal}^*\cdot p)e^{-\phi-12A}
 \\
 -&3c_0 {\mathcal{A}}e^{\phi/4 - 4A}
+e^{6A}\mathcal{B}
+\tfrac12\Big(
{\mathcal{A}}e^{3A}+(3c_0\chi -\Xi)e^{-\phi/4 - 9A}
\Big)^2
~.
}}
The ten-dimensional $H$-field equation of motion, cf.~\eqref{beomf3}, implies, 
\eq{\label{12}
\d\left(
e^{-\phi+12A}\star_4 \d\beta\right)
= 3c_0(\d\gamma - \alpha\wedge \d\chi)- D\xi'_A\wedge D\xi^A + e^{\phi/2+18A} \varphi \d\alpha+e^{\phi/4+12A}\mathcal{A}\d\alpha
~.}
This is consistent with the  equation  of motion for $\beta$ coming from \eqref{blipuy}, as can be seen by eliminating $\varphi$ using \eqref{phisol}. 
Eq.~\eqref{12} can thus be  solved to trade the two-form $\beta$ for the axion $b$,
\eq{\label{prhd2qr}
 \d \beta =e^{\phi-12A}\star_4\left[\d b+\tfrac12(\xi'_A\d\xi^A-\xi^A\d\xi'_A) + 3c_0(\gamma - \chi \alpha)+\Xi\alpha
\right]
~,}
where we have taken \eqref{covxi} into account.  
Using \eqref{prhd2qr}, we obtain the action in terms of the axion,
\boxedeq{\spl{\label{ctr2x}
S_4=\int&\d^4 x\sqrt{g}
\Big(
R
 - 24 (\partial A)^2 
 -\tfrac{1}{2} (\partial \phi)^2  -2g_{a\bar{b}}\d z^a\d\bar{z}^b
-\tfrac{3}{2}  e^{-4A - \phi}(\partial \chi)^2\\
&+ \tfrac{1}{2} e^{-6A + \phi/2} (D\xi'+\mcal \cdot D\xi)\cdot\Im\mcal^{-1}\cdot(D\xi'+{\mcal}^*\cdot D\xi)\\
&
-\tfrac{1}{4} e^{3\phi/2 + 6A} \d\alpha^2
-\tfrac{3}{4}  e^{\phi/2 + 2A} (\d\gamma - \alpha\swed\d\chi)^2
-\tfrac{1}{2} e^{\phi - 12A} (D b )^2-V
 \Big)
%CS terms:
+\int 
3 \chi\ \d\gamma\swed\d\gamma  
~,
}}
where we have defined,   
\eq{\spl{\label{43}
Db:=\d b+\tfrac12(\xi'_A\d\xi^A-\xi^A\d\xi'_A) + 3c_0(\gamma - \chi \alpha)+\Xi\alpha
~.} }
Note that,  for $c_0\neq0$, \eqref{58x}-\eqref{43} remain invariant under the 
shift \eqref{shiftui}.

\subsection{Comparison with $\mathcal{N}=2$, 4d supergravity}

In terms of the  fields introduced in \eqref{eqti}, \eqref{svecs},  and the K\"{a}hler potential \eqref{kpot1i}, 
the action    \eqref{ctr2x}  takes the following form,
\boxedeq{\spl{\label{actionfinal2}
S_4=\int&\d^4 x\sqrt{g_4}
\Big(
R
 - 2\partial_t\partial_{\bar{t}}K_V|\d t|^2-2g_{a\bar{b}}\d z^a\d\bar{z}^b
-2(\d\hat{\phi})^2- \tfrac{1}{2} e^{4\hat{\phi}} (D b)^2
 \\
&+ \tfrac{1}{2} e^{2\hat{\phi}}  (D\xi'+\mcal \cdot D\xi)\cdot\Im\mcal^{-1}\cdot(D\xi'+{\mcal}^*\cdot D\xi)
\\
&-\tfrac{1}{4} v^{3} \mathcal{F}^2
-\tfrac{3}{4}  v (\tilde{\mathcal{F}}+\chi \mathcal{F})^2
-V
 \Big) 
+\int 
3 \chi \tilde{\mathcal{F}}\swed\tilde{\mathcal{F}}
+ \chi^3  {\mathcal{F}}\swed {\mathcal{F}}
+3\chi^2  {\mathcal{F}}\swed\tilde{\mathcal{F}}
~.
}}
It can be verified that 
  \eqref{actionfinal2} is of the form of $\mathcal{N}=2$ supergravity, given in \eqref{n2}. 
For the vector multiplet sector this was already shown in section \ref{sec:universal}. 
For the hypermultiplet sector we proceed as follows:  
the  Killing vectors can  be read off by comparing \eqref{covxi}, \eqref{43} and \eqref{43prime},
\eq{\label{kilac}
k^u_{I=0}\partial_{q^u}=-\Xi\partial_{b}-q_A\partial_{\xi'_A}-p^A\partial_{\xi^A}~;~~~
k^u_{I=1}\partial_{q^u}=-3c_0\partial_{b}
~,}
where $u,v=1,\dots, 4n_H$, and the  $q^u$ coordinates are identified with $\hat{\phi}$, $b$,  $\xi^A$, $\xi'_A$, $\Re z^a$, $\Im z^a$.  
The hypermultiplet kinetic terms are manifestly of the form dictated by  the quaternionic metric \eqref{qmetric}. 
Finally, to show that the potential \eqref{58x} is of the from 
given in \eqref{potac}, we compute,
\eq{\spl{\label{32}
h_{uv}k_I^uk_J^v=\frac12 \left(
\begin{matrix}
   e^{4\hat{\phi}}\Xi^2- e^{2\hat{\phi}}(q+\mcal\cdot p)\cdot\Im\mcal^{-1}\cdot(q+{\mcal}^*\cdot p)   & -3c_0e^{4\hat{\phi}}\Xi \\
   -3c_0e^{4\hat{\phi}}\Xi   & 9c_0^2e^{4\hat{\phi}}
\end{matrix}\right)
~,}}
while \eqref{32bi}, \eqref{33i} remain unchanged. 
Substituting the latter together with  \eqref{32} into \eqref{potac} reproduces precisely the potential given in \eqref{58x}, in the absence of condensates: $\mathcal{A}=\mathcal{B}=0$.

\section{Positivity of the quartic gravitino condensate}\label{sec:positivity}

On ALE spaces one has the possibility to choose  a gauge in which the spin connection is self-dual \cite{Eguchi:1980jx}. 
Covariantly-constant negative-chirality spinors in this gauge  are just constant, and 
we may choose their basis  as follows,   
\eq{\label{ccs}
\theta^{(1)}=
  \left( {\begin{array}{c}
  1  \\
  0 
  \end{array} } \right)~;~~~
  \theta^{(2)} =
  \left( {\begin{array}{c}
  0  \\
  1 
  \end{array} } \right)
~.}
Let us now consider a spin-1 field, i.e. a field transforming in the three-dimensional irreducible representation of the $su(2)$ algebra. This can be represented as a field with 
two symmetric spinor indices of the same chirality, $\phi_{\alpha\beta}=\phi_{\beta\alpha}$ (positive chirality) or $\phi^{\alpha\beta}=\phi^{\beta\alpha}$ (negative chirality). 
The Atiyah-Patodi-Singer theorem for a spin-1 field on an ALE space predicts that the number of positive-chirality zeromodes of the Dirac operator minus the number of  negative-chirality zeromodes is equal to  the Hirzebruch signature $(\tau)$ of the ALE space. Moreover it can be seen that there are no renormalizable negative-chirality zeromodes. It then follows that there are exactly $\tau$ spin-1 zeromodes of positive chirality.

By a similar argument as before, on ALE spaces there are exactly $2\tau$ spin-3/2 (gravitino) zeromodes  of positive chirality. 
They   can be constructed as follows \cite{Hawking:1979zs},
\eq{\label{psiconstr}
\psi^{(iI)}_{\mu\alpha}=\phi^{(I)}_{\alpha\beta}\left(\tilde{\theta}^{(i)}\gamma_\mu\right)^\beta
~;~~~i=1,2~;~I=1,\dots,\tau
~,}
where $\theta^{(i)}$ are  covariantly-constant spinors of negative chirality, cf.~\eqref{ccs}, and $\phi^{(I)}_{\alpha\beta}$ are  positive-chirality spin-1 zeromodes.

Note that in this gauge the $\psi^{(iI)}_{\mu}$'s are automatically traceless, $\gamma^\mu\psi^{(iI)}_{\mu}=0$, as follows from,
\eq{\label{mid}
\left(C^{-1}\gamma^\mu\right)_{\gamma}{}^{(\alpha}
\left(C^{-1}\gamma_\mu\right)_{\delta}{}^{\beta)}=0~.} 
Indeed 
if the above were not true, given an arbitrary spin-1 field of positive chirality $\phi$,  
one could construct a spin-1 field of negative chirality $\phi'$, via 
$\phi^{\prime\gamma\delta}=(\gamma_\mu)^{\alpha\gamma}(\gamma^\mu)^{\beta\delta} \phi_{\alpha\beta}$. 
This is a contradiction, since the positive- and negative-chirality spin-1 representations transform under the $({\bf 3},{\bf 1})$ and $({\bf 1},{\bf 3})$ of $so(4)\cong su(2)\oplus su(2)$, respectively.

In  the path-integral over metrics approach, the leading contribution to the quartic-fermion condensate would come from 
spaces with four gravitino zero modes ($\tau=2$). 
To a first approximation in the gravitational coupling, 
and up to a positive proportionality constant, 
it is obtained from the quartic terms in  \eqref{calbdef}, by replacing 
each term by its expansion on the basis of zero modes \eqref{psiconstr},
\eq{
\psi_{\mu_1}\psi_{\mu_2}\psi'_{\mu_3}\psi'_{\mu_4}\rightarrow \sum_{i_k, I_l=1}^2 \psi^{(i_1I_1)}_{\mu_1}\psi^{(i_2I_2)}_{\mu_2}\psi^{(i_3I_3)}_{\mu_3}\psi^{(i_4I_4)}_{\mu_4}
~,} 
where  the sum is over all antisymmetrized permutations of the zero modes \cite{Hawking:1979zs,Konishi:1988mb}. 

We will now show that,
\eq{
\tilde{\psi}^{(iI)}_{[\mu}\psi^{(jJ)}_{\nu]}- \tilde{\psi}^{(jJ)}_{[\mu}\psi^{(iI)}_{\nu]}=0
~,}
which implies that, within the present approximation,  the first two terms on the right-hand side of the second line of \eqref{calbdef} do not contribute to the quartic condensate, leaving as the only 
contribution the last positive-definite term. Indeed, using the form of the zero modes eq.~\eqref{psiconstr} and the symmetry properties \eqref{symprp}, we find,
\eq{
\tilde{\psi}^{(iI)}_{[\mu}\psi^{(jJ)}_{\nu]} =
\left(\tilde{\theta}^{(i)}\gamma_{[\mu}\right)^\alpha \left(\tilde{\theta}^{(j)}\gamma_{\nu]}\right)^\beta (\phi^{(I)}\cdot C^{-1}\phi^{(J)})_{\alpha\beta}
= \tilde{\psi}^{(jJ)}_{[\mu}\psi^{(iI)}_{\nu]}
~.}
Finally one can show that 
the   last term in \eqref{calbdef} gives,
\eq{\label{cse}
\mathcal{B}\propto
\text{tr}\left((M^{(1)})^2\right)\text{tr}\left((M^{(2)})^2\right)-\left(\text{tr}\left(M^{(1)}\cdot M^{(2)}\right)\right)^2\geq 0
~,}
where we have defined the matrix $(M^{(I)})^\alpha{}_\beta:=C^{\alpha\delta}\phi^{(I)}_{\delta\beta}$, and we have taken into account \eqref{mid}, \eqref{a4}, \eqref{fgt}, and the idenitites 
$\tilde{\theta}^{(1)}{\theta}^{(2)}=1$,   $\tilde{\theta}^{(1)}{\theta}^{(1)}=\tilde{\theta}^{(2)}{\theta}^{(2)}=0$. Expression \eqref{cse} is non-negative  by 
virtue of the Cauchy-Schwarz inequality.

\section{NK universal consistent truncation}\label{sec:nk}

In this section we extend the consistent truncation of \cite{Terrisse:2018qjm} to include the vectors $\alpha$, $\gamma$, the two-form $\beta$ and the additional scalar $\xi'$, 
to complete the bosonic content of the universal  sector of one gravity multiplet, one vector multiplet and one hypermultiplet of $\mathcal{N}=2$ 4d supergravity.\footnote{The scalar 
$\xi$ of the present paper was previously deonted by $\gamma$ in \cite{Terrisse:2018qjm}.}  
Explicitly we set,
\eq{\spl{\label{foranscy0}
F&=\d\alpha+m\beta+m\chi J~;~~~ H=\d\chi \swed J-6\omega \chi \text{Re}\Omega+\d\beta\\
G&=\varphi\text{vol}_4+\frac12  (m\chi^2 + \xi)J\swed J+ J\swed (\d\gamma - \alpha\wedge \d\chi+m\chi\beta)
 -\frac{1}{8\omega}\d \xi\swed\text{Im}\Omega
-\frac{1}{2}D\xi'\swed\text{Re}\Omega
~,}}
where $\xi(x)$, $\varphi(x)$ are real 4d scalars. The covariant derivative is given by,
\eq{
D\xi':=\d\xi'+12\omega(\gamma-\chi\alpha)
~.}
We have chosen to express $H$ in terms of the 4d potential $\beta$ instead of the axion. 

In the massive case, $m\neq0$, the one-form $\alpha$ can be absorbed in (``eaten by'') the two-form $\beta$ via the St\"{u}ckelberg mechanism. This is however  no longer true in the 
massless limit. In order to be able to treat both massive and massless cases on an equal footing, we keep both $\alpha$ and $\beta$ in the form ansatz. 
The massless limit  is obtained by simply setting $m= 0$ in  \eqref{foranscy0}. 

 Taking into account that for a  NK manifold we have,  
\eq{\spl{
\d J&=-6\omega\mathrm{Re}\Omega \\
\d\mathrm{Im}\Omega&= 4\omega J\wedge J ~,
\label{torsionclassesbnk}
}}
the ansatz \eqref{foranscy0} can be seen to automatically satisfy the Bianchi identities (\ref{bi}).

Our ansatz for the ten-dimensional metric reads,
\eq{\label{tdmank}\d s^2_{(10)} =e^{2A(x)}\left(e^{2B(x)} g_{\mu\nu}\d x^{\mu}\d x^{\nu}+g_{mn}\d y^m\d y^n 
\right)~,
}
where the scalars $A$, $B$ only depend on the four-dimensional coordinates $x^\mu$. 
This gives,
\eq{\spl{
F^2_{mn}&= m^2\chi^2e^{-2A}g_{mn}~;~~~ F^2_{\mu\nu} = e^{-2A-2B} (\d\alpha+m\beta)^2_{\mu\nu} \\
F^2 &= e^{-4A-4B} (\d\alpha+m\beta)^2 +6m^2\chi^2e^{-4A}\\
H^2_{mn}&= 2e^{-4A-2B}(\partial\chi)^2g_{mn} +144e^{-4A}\omega^2\chi^2 g_{mn}~;~~~
H^2_{\mu\nu} = 6e^{-4A}\partial_{\mu}\chi\partial_{\nu}\chi+e^{-4A-4B}h^2_{\mu\nu}\\
H^2 &= 18e^{-6A-2B}(\partial\chi)^2+e^{-6A-6B}h^2+864e^{-6A}\omega^2\chi^2\\
G^2_{mn} &= 3e^{-6A-2B}\Big[\tfrac{1}{16\omega^2}(\partial \xi)^2+(D\xi')^2\Big]g_{mn}+12e^{-6A}(m\chi^2+\xi)^2g_{mn}\\
&+3e^{-6A-4B}(\d\gamma - \alpha\wedge \d\chi+m\chi\beta)^2 g_{mn} \\
G^2_{\mu\nu} &=-6e^{-6A-6B}\varphi^2g_{\mu\nu}  + 6e^{-6A}( \tfrac{1}{16\omega^2}\partial_{\mu}\xi\partial_{\nu}\xi+D_{\mu}\xi' D_{\nu}\xi')\\
&+18e^{-6A-2B}(\d\gamma 
- \alpha\wedge \d\chi  +m\chi\beta)^2_{\mu\nu}\\
G^2 &= -24 e^{-8A-8B}\varphi^2
+24e^{-8A-2B}\Big[\tfrac{1}{16\omega^2}(\partial \xi)^2+(D\xi')^2\Big]+72(m\chi^2+\xi)^2e^{-8A}\\
&+36e^{-8A-4B}(\d\gamma - \alpha\wedge \d\chi +m\chi\beta)^2
~,}}
where the contractions on the left-hand sides above are computed with respect to the ten-dimensional metric; the contractions 
on the right-hand sides are taken with respect to the unwarped metric. It is also useful to note the following expressions,
\eq{\spl{\label{hodsr}
\star_{10} F &= \frac{1}{6}e^{6A} \star_4(\d\alpha+m\beta)\swed J^3 + \frac12 m\chi e^{6A+4B} \text{vol}_4\swed J^2
\\	
\star_{10} H &=  \tfrac12 e^{4A+2B} \star_{4}\!\d\chi\swed   J^2
+ \tfrac16 e^{4A-2B} \star_{4}\!h\swed   J^3+6\omega\chi e^{4A+4B}\text{vol}_4\swed \text{Im}\Omega
\\
\star_{10} G &= -\tfrac16 \varphi e^{2A-4B}  J^3
+(m\chi^2+\xi) e^{2A+4B} \text{vol}_4\swed J+\tfrac12 e^{2A}\star_{4}(\d\gamma - \alpha\wedge \d\chi+m\chi\beta)\swed J^2\\
&~~~~\!+\tfrac{1}{8\omega} e^{2A+2B}   \star_{4}\!\d \xi\swed  \text{Re}\Omega
-\tfrac{1}{2} e^{2A+2B}   \star_{4}\! D\xi'\swed  \text{Im}\Omega
~,}}
where the four-dimensional Hodge-star is taken with respect to the unwarped metric.

Plugging the above ansatz into the ten-dimensional EOM \eqref{beomf1}-\eqref{beomf3} we obtain the following: the internal $(m,n)$-components of the Einstein 
equations  read,
\eq{\spl{\label{et1}
20e^{2B}\omega^2&=e^{-8A-2B}\nabla^{\mu}\left(
e^{8A+2B}\partial_{\mu}A
\right)+\frac{1}{16}m^2e^{5\phi/2+2A+2B}
+\frac{5}{16}e^{3\phi/2-2A+2B}m^2\chi^2\\
&-\frac{1}{32} e^{3\phi/2-2A-2B} (\d\alpha+m\beta)^2
+\frac18e^{-\phi-4A}(\partial\chi)^2
-\frac{1}{48}e^{-\phi-4A-4B}h^2+18 e^{-\phi-4A+2B}\omega^2\chi^2\\
&-\frac{1}{32}e^{\phi/2-6A-2B}(\d\gamma - \alpha\wedge \d\chi+m\chi\beta)^2
+\frac{1}{16}e^{\phi/2-6A}
\Big[\tfrac{1}{16\omega^2}(\partial \xi)^2+(D\xi')^2\Big]
\\
&+\frac{3}{16} 
e^{\phi/2-6A-6B}\varphi^2
+\frac{7}{16}e^{\phi/2-6A+2B} (m\chi^2+\xi)^2
~.}}

The external $(\mu,\nu)$-components read,
\eq{\spl{\label{et2}
R^{(4)}_{\mu\nu}&=
g_{\mu\nu}\left(\nabla^{2}A+\nabla^{2} B+
8(\partial A)^2+2(\partial B)^2+10\partial A\cdot \partial B\right)
\\
&-8\partial_{\mu}A\partial_{\nu}A-2\partial_{\mu}B\partial_{\nu}B
-16\partial_{(\mu}A\partial_{\nu)}B+8\nabla_{\mu}\partial_{\nu}A+2\nabla_{\mu}\partial_{\nu}B
\\
&+\frac32 e^{-\phi-4A} \partial_{\mu}\chi\partial_{\nu}\chi
+\frac12 e^{3\phi/2 -2A-2B} (\d\alpha+m\beta)^2_{\mu\nu}
+\frac14 e^{\phi-4A-4B} h^2_{\mu\nu}
+\frac12 \partial_{\mu}\phi\partial_{\nu}\phi
\\
&+\frac{1}{2} e^{\phi/2-6A} ( \tfrac{1}{16\omega^2}\partial_{\mu}\xi\partial_{\nu}\xi+D_{\mu}\xi' D_{\nu}\xi')
+\frac{3}{2} e^{\phi/2-6A-2B}(\d\gamma - \alpha\wedge \d\chi+m\chi\beta)^2_{\mu\nu}
 \\
&+\frac{1}{16} g_{\mu\nu}\Big(  
- \frac{1}{2} e^{3\phi/2-2A-2B} (\d\alpha+m\beta)^2
-\frac{1}{3}e^{\phi-4A-4B}h^2
\\
&-3e^{\phi/2-6A} \Big[\tfrac{1}{16\omega^2}(\partial \xi)^2+(D\xi')^2\Big]-6e^{-\phi-4A}(\partial\chi)^2
-5e^{\phi/2-6A-6B}\varphi^2 
\\
&+ m^2e^{5\phi/2+2A+2B}
-3m^2\chi^2 e^{3\phi/2-2A+2B} 
-288e^{-\phi-4A+2B}\omega^2\chi^2\\
&-9 (m\chi^2+\xi)^2e^{\phi/2-6A+2B} -\frac{9}{2}e^{\phi/2-6A-2B}(\d\gamma - \alpha\wedge \d\chi+m\chi\beta)^2
\Big)
~,}}
while the mixed $(\mu,m)$-components are automatically satisfied. 
The dilaton equation reads,
\eq{\spl{\label{et3}
0&=e^{-10A-4B}\nabla^{\mu}\left(
e^{8A+2B}\partial_{\mu}\phi
\right)
-\frac{1}{4}e^{\phi/2-8A-2B}\Big[\tfrac{1}{16\omega^2}(\partial \xi)^2+(D\xi')^2\Big]
\\
&-\frac{3}{8}e^{3\phi/2-4A-4B} (\d\alpha+m\beta)^2
-\frac{5}{4}m^2e^{5\phi/2}
-\frac{9}{4}e^{3\phi/2-4A}m^2\chi^2
\\
&+\frac32e^{-\phi-6A-2B}(\partial\chi)^2
+\frac{1}{12}e^{-\phi-6A-6B}h^2+72 e^{-\phi-6A}\omega^2\chi^2\\
&+\frac{1}{4}
e^{\phi/2-8A-8B}\varphi^2-\frac{3}{4}(m\chi^2+\xi)^2 e^{\phi/2-8A}-\frac{3}{8}e^{\phi/2-8A-4B}(\d\gamma - \alpha\wedge \d\chi+m\chi\beta)^2
~.}}
The $F$-form equation of motion 
reduces to the condition,
\eq{\spl{\label{11}
\d\left(e^{3\phi/2+6A} \star_4 (\d\alpha+m\beta)\right) &= \varphi e^{\phi/2+2A-4B}\d\beta - 3e^{\phi/2+2A}\d\chi\wedge\star_4(\d\gamma - \alpha\wedge \d\chi+m\chi\beta)\\
&-12\omega\chi e^{\phi/2+2A+2B}\star_4  D\xi'
~.}}
The $H$-form equation reduces to the following two equations,
\eq{\spl{\label{hfeom}
\d\left(
e^{-\phi+4A+2B}\star_4\d\chi
\right)  &= \Big[(m\chi^2+\xi)\varphi 
-48\omega^2 e^{-\phi+4A+4B}\chi
-2me^{\phi/2+2A+4B}( m\chi^2+\xi)\chi\\
&
- e^{3\phi/2+6A+4B}m^2\chi
\Big]\text{vol}_4 -e^{\phi/2+2A} (\d\alpha+m\beta)\wedge\star_4(\d\gamma - \alpha\wedge \d\chi+m\chi\beta)
\\
&+ (\d\gamma - \alpha\wedge \d\chi+m\chi\beta)\wedge(\d\gamma - \alpha\wedge \d\chi+m\chi\beta)
~,}}
and, 
\eq{\spl{\label{seqh}
\d\left(
e^{-\phi+4A-2B}\star_4 \d\beta\right)
&= 3( m\chi^2+\xi)(\d\gamma - \alpha\wedge \d\chi+m\chi\beta)- \frac{1}{4\omega}\d\xi\wedge D\xi' + e^{\phi/2+2A-4B} \varphi (\d\alpha+m\beta)\\
&-e^{3\phi/2+6A}m\star_4(\d\alpha+m\beta)
-3e^{\phi/2+2A}m\chi \star_4 (\d\gamma - \alpha\wedge \d\chi+m\chi\beta)
~.}}
The $G$-form equation of motion reduces to,
\eq{\spl{
\label{gfeom1}
\d\left(e^{\phi/2+2A+2B}\star_4\d\xi\right)  &= 4\omega h\swed D\xi'
-48\omega^2 \left(  e^{\phi/2+2A+4B}( m\chi^2+\xi) -\chi\varphi\right)\text{vol}_4
\\
\d\left(e^{\phi/2+2A+2B}\star_4 D\xi'\right) &= -\frac{1}{4\omega}h\swed\d\xi\\
\d\left(e^{\phi/2+2A} \star_4(\d\gamma - \alpha\wedge \d\chi+m\chi\beta)\right) &=  2\d\chi\swed
(\d\gamma+m\chi\beta)+( m\chi^2+\xi)\d\beta
+4\omega e^{\phi/2+2A+2B}  \star_4 D\xi'
~,}}
together with the constraint,
\eq{\label{gfeom2}
0=\d\left(
 \varphi e^{\phi/2+2A-4B}+ m\chi^3+ 3\xi\chi
\right)
~.}
The latter can be  integrated to solve for $\varphi$ in terms of the other fields,
\eq{\label{fg}
\varphi=\left(C-m\chi^3-3\xi\chi\right)e^{-2A+4B-\phi/2}
~,}
where $C$ is an arbitrary constant.

\subsection{Action}

Setting $B=-4A$, it can be seen that  the  dilaton and Einstein equations \eqref{et1}-\eqref{et3}, can all be integrated 
to the following 4d action,\footnote{To obtain \eqref{21},  the value of $\varphi$, as given in \eqref{fg}, must be substituted in \eqref{et3} 
{\it before} integrating with respect to the dilaton.
This action was first obtained in \cite{KashaniPoor:2007tr}  by directly inserting the form ansatz into the 10d IIA theory, a procedure that does not in general lead 
to a CT. However it was shown that certain susy solutions of the 4d theory uplift to a well-known class of solutions \cite{Behrndt:2004km,Lust:2004ig}. 
Ref.~\cite{KashaniPoor:2007tr} also shows that the 4d action is consistent with $\mathcal{N}=2$ gauged supergravity in its  formulation with (massive) tensors \cite{Theis:2003jj,DallAgata:2003sjo,DAuria:2004yjt}.
}
\eq{\spl{\label{21}
S_4= &\int\d^4 x\sqrt{g}
\Big(
R
 - 24 (\partial A)^2 
 -\tfrac{1}{2} (\partial \phi)^2  
-\tfrac{3}{2}  e^{-4A - \phi}(\partial \chi)^2 
- \tfrac{1}{2} e^{-6A + \phi/2} \left[\tfrac{1}{16\omega^2}(\partial \xi)^2+(D \xi')^2\right]\\
&-\tfrac{1}{4} e^{3\phi/2 + 6A} (\d\alpha+m\beta)^2
-\tfrac{3}{4}  e^{\phi/2 + 2A} (\d\gamma - \alpha\swed\d\chi+m\chi\beta)^2
-\tfrac{1}{12} e^{-\phi + 12A} \d\beta^2\\
&-\tfrac12 m^2e^{-6A + 5\phi/2} 
-\tfrac32 m^2\chi^2e^{-10A + 3\phi/2} 
-72 \omega^2\chi^2e^{-12A -\phi} 
-\tfrac32 e^{-14A + \phi/2}  (\xi+m\chi^2)^2\\
&-\tfrac{1}{2}  e^{-\phi/2 - 18A} 
\left(C-m\chi^3-3\xi\chi\right)^2
 +120\omega^2e^{-8A}\Big)
+S^{\text{SC}}
~.}}
The Chern-Simons terms above, $S^{\text{SC}}$, cannot be determined from  \eqref{et1}-\eqref{et3}: 
 they must be reconstituted from the form equations of motion -- which at the same time impose several additional 
 consistency conditions on the action \eqref{21}.   We thus  find that the form equations of motion are consistent with the 
Chern-Simons term, 
\eq{\spl{
S^{\text{SC}}=
\int &\beta\swed\left(
-\tfrac{1}{4\omega}\d\xi\swed\d\xi'
+3m\chi^2 \d\gamma
+\d\left[ (C-m\chi^3-3\xi\chi)\alpha+3\xi \gamma)\right]
\right)\\
&+ m(\tfrac12 C+m\chi^3)\beta\wedge\beta
+3 \chi\ \d\gamma\swed\d\gamma 
~.
}}

\subsection{Dualization}

To dualize the antisymmetric two-form potential $\beta$ to an axion $b$, 
which can only be performed in the massless limit $(m=0)$, we proceed as follows: 
first note that the second equation in \eqref{gfeom1} can be solved in order to express  $D\xi'$ in terms of a dual two-form $\tilde{\xi}'$,
\eq{\label{17}
e^{\phi/2-6A}\star_4 D\xi' = -\frac{1}{4\omega}h \xi+\d\tilde{\xi}'~.}
Using the above, the last equation in \eqref{gfeom1} can be integrated to,
\eq{\label{18}
 e^{\phi/2+2A} \star_4(\d\gamma - \alpha\wedge \d\chi)=  2\d\chi\swed \gamma+ 4\omega\tilde{\xi}'+\d\tilde{\gamma}
 ~, }
in terms of a dual one-form $\tilde{\gamma}$. Eq.~\eqref{11}   then   integrates to,
\eq{ \label{19}
e^{3\phi/2+6A} \star_4 \d\alpha =C\beta-12\omega\chi\tilde{\xi}'-3\d\chi\swed\tilde{\gamma}
~,}
where we have also taken \eqref{fg} into account.  
Substituting \eqref{17}-\eqref{19} into \eqref{seqh}, we obtain the desired dualization condition,
\eq{\label{seqhb}
e^{-\phi+12A}\star_4 \d\beta 
=\d b+\mathfrak{A}~,
}
where we introduced the   one-form,
\eq{\label{cofd}
\mathfrak{A}:=
\tfrac{1}{8\omega}\left(\xi\d\xi'-\xi'\d\xi \right)+
 (C-3\xi\chi ) \alpha +3\xi\gamma
~.}
Note that the dependence on $\tilde{\gamma}$, $\tilde{\xi}'$ drops out of  the  $\beta$-dualization expression \eqref{seqhb}. In terms of the axion, the action reads,
\boxedeq{\spl{\label{21b}
S_4= &\int\d^4 x\sqrt{g}
\Big(
R
 - 24 (\partial A)^2 
 -\tfrac{1}{2} (\partial \phi)^2  
-\tfrac{3}{2}  e^{-4A - \phi}(\partial \chi)^2 
- \tfrac{1}{2} e^{-6A + \phi/2} \left[\tfrac{1}{16\omega^2}(\partial \xi)^2+(D \xi')^2\right]\\
&-\tfrac{1}{4} e^{3\phi/2 + 6A} (\d\alpha)^2
-\tfrac{3}{4}  e^{\phi/2 + 2A} (\d\gamma - \alpha\swed\d\chi)^2
-\tfrac{1}{12} e^{\phi - 12A} (\d b+\mathfrak{A})^2\\
& 
-72 \omega^2\chi^2e^{-12A -\phi} 
-\tfrac32 e^{-14A + \phi/2}  \xi^2
-\tfrac{1}{2}  e^{-\phi/2 - 18A} 
\left(C-3\xi\chi\right)^2
 +120\omega^2e^{-8A}\Big)
\\
&
+
\int 
3 \chi\ \d\gamma\swed\d\gamma 
~.}}
One can also make contact with the CY reduction of section \ref{sec:universal} in the absence of condensates, 
by   setting, 
\eq{\label{73}
\chi\rightarrow\chi-\tfrac{1}{12\omega}b_2~;~~~\xi\rightarrow c_0+4\omega\xi
~;~~~
b\rightarrow b -\tfrac{1}{8\omega}c_0\xi'
~;~~~
C\rightarrow  -\tfrac{1}{4\omega}b_2c_0
~,  
}
and then taking the limit of vanishing torsion class: 
$\omega\rightarrow 0$.\footnote{Compared to section \ref{sec:universal}, \eqref{73} corresponds to having a real background three-form flux  $b_0$ (i.e.~setting $b_1=0$). This  can easily be achieved by shifting the phase of the CY three-form  $\Omega$ by a suitable constant.}

\section{Discussion}\label{sec:conclusions}

One of the motivations of constructing consistent truncations in the presence of condensates, is that it gives us a way of fixing the dependence of the condensates on the moduli. 
In the case considered here, the quadratic and quartic condensates decompose into a 4d  moduli-independent piece (which we denoted by $\mathcal{A}$ and $\mathcal{B}$ respectively), and a moduli-dependent factor. For example, in the case  of IIA compactified on a CY with $h^{2,1}=0$, 
the quadratic condensate is constrained by the consistency of the theory to enter the potential in the combination  $v^{3/4}e^{-3\hat{\phi}/2}\mathcal{B}$, cf.~\eqref{58}, \eqref{eqti}. I.e. imposing the consistency of the truncation determines the dependence of the condensates on the volume modulus and the 4d dilaton.

It can be easily established that the effective theory of IIA compactified on a CY with $h^{2,1}\geq 1$ admits formal de Sitter solutions supported by a positive quartic gravitino condensate, just as in \cite{Terrisse:2019usq}. 
As in that reference, there are  flat directions  for  the axion $b$ and the RR axions $\xi^A$, $\xi_A'$, all of which are compact scalars. 
These  solutions require a strictly positive quartic condensate, $\mathcal{B}>0$, but leave the quadratic condensate $\mathcal{A}$ unconstrained. 
In the presence of four spin-3/2 zero modes, a quartic condensate will be induced at one-loop order in the 4d gravitational coupling. 
In the presence of two spin-3/2 zero modes, in which case $\mathcal{A}\neq0$, $\mathcal{B}=0$ at one loop, a positive quartic condensate will plausibly be induced 
in the effective action at two loops in the gravitational coupling, $\mathcal{B}\propto\mathcal{A}^2$. 

As we have seen in section \ref{sec:positivity},   $\mathcal{B}$ is non-negative  in the background of ALE  gravitational instantons with four spin-3/2 zero modes.  
 However it is not clear whether such a condensate is non-zero. Imposing that the noncompact space should asymptote  $\mathbb{R}^4$ at infinity, would exclude ALE gravitational instantons altogether: for the  ALE instantons to contribute to the path integral, we must accept spacetimes with non-standard asymptotics. It is not clear whether this can make sense physically,\footnote{In the case of the Eguchi-Hanson instanton, the discrete identifications are given by $(t,\vec{x})\rightarrow-(t,\vec{x})$ at asymptotic infinity. To avoid closed timelike loops, upon Wick-rotation to Lorentzian signature, it would suffice to restrict to $t>0$, at the cost of rendering the space incomplete.} although the fact that the Eguchi-Hanson ALE instanton \cite{Eguchi:1978gw} can be thought of as  T-dual to a string theory solution with $\mathbb{R}^4$  asymptotics, would appear to support the admissibility of ALE instantons \cite{Bianchi:1994gi}.

If we allow spacetimes with ALE asymptotics, for the de Sitter solutions of \cite{Terrisse:2019usq} and the present paper to go through, one needs not only non-vanishing  condensates, but also non-vanishing background fluxes: $|b_0|, c_0\neq 0$. However, it has not been shown that ALE instanons can be embedded in  a  4d theory with such background fluxes. One way to address both of these issues would be to construct other gravitational instantons with  $\mathbb{R}^4$ or AdS$_4$ asymptotics\footnote{As already mentioned, our derivation of the 4d action \eqref{ctr2}, \eqref{blipuy} is valid for  condensates  induced by any  gravitational instantons with 4d gravitino zero modes of positive chirality. In particular our results should  also be applicable to the case of K3 instantons glued into asymptotic $\mathbb{R}^4$, as considered in \cite{Hebecker:2019vyf}.} 
in the presence of background fluxes.~Such backgrounds would have non-vanishing profiles for the matter fields.

A CT may  truncate out some of the ``light'' degrees of freedom. As a result, solutions 
obtained within the CT may be unstable when uplifted to the higher-dimensional theory.  
This phenomenon  would not occur if the universal CT obtained in section \ref{sec:universal} comes from compactification on a CY with $h^{1,1}=1, h^{2,1}=0$, since in this case all light degrees of freedom are already present in the CT. However, there is no known CY with such Hodge numbers to date, although there is no known theorem precluding its existence either.\footnote{Note that such a CY, if it exists,  cannot have a mirror.}   
On the other hand, the effective action of section \ref{sec:cycomp} should adequately capture all light degrees of freedom (at least below the energy scales set by  the 
compactification and the background flux), but need not be a CT.

Naively relying on the 4d effective action, in particular if the uplift is unknown, may lead to miss some crucial physics, see  \cite{Font:2019uva} for a recent example.~A more important objection  states that one  should reject   altogether solutions of the 4d effective theory, such as de Sitter space, whose asymptotics are 
drastically different from 
the ones (flat Minkowski) used in deriving the 10d low-energy string effective action, see 
\cite{Banks:2019oiz} for a recent discussion.~In other words, different asymptotic spacetime conditions should correspond to different quantum gravity models, 
and there is no a priori justification to using the low-energy effective action to extrapolate from one set of asymptotics to another. If this is true, string theory  would 
currently have little to say about spacetimes with asymptotics other than flat or AdS space.

\section*{Acknowledgment}

The present paper is an extended version of my talk at CORFU2019. 
I am grateful to  the organizers of the conference for providing a pleasant and stimulating environment. 
I am also grateful to the physics department of TU Wien for kind hospitality. 
I would like to thank David Andriot, Niccol\`{o} Cribiori, Thomas Van Riet, Stefan Theisen and Timm Wrase for valuable discussions.

\appendix

\section{Conventions}\label{app:spin}

We are following 
the ``superspace conventions'' where all  forms are given by,
\eq{
\Phi_{(p)}=\frac{1}{p!}\Phi_{m_1\dots m_p}\d x^{m_p}\swed\dots\swed\d x^{m_1}~;~~~
\d\Big( \Phi_{(p)} \swed\Psi_{(q)}\Big)=\Phi_{(p)} \swed\d\Psi_{(q)}
+(-1)^q\d\Phi_{(p)} \swed\Psi_{(q)}~.
}
The Hodge star of a $p$-form in $d$ dimensions is given by,
\eq{
\star (\d x^{a_1}\wedge\dots\wedge \d x^{a_p})=\frac{1}{(d-p)!}\varepsilon^{a_1\dots a_p}{}_{b_1\dots b_{d-p}} \d x^{b_1}\wedge\dots\wedge \d x^{b_{d-p}}
~,}
so that for any spacetime signature,
\eq{
\Phi_{(p)}\wedge\star\Phi_{(p)}
=\frac{1}{p!}\Phi^2\text{vol}_d 
~,}
where we have defined,
\eq{
\text{vol}_d:= \star1~;~~~\Phi^2:=\Phi_{m_1\dots m_p} \Phi^{m_1\dots m_p}
~.}
The following expressions are also useful,
\eq{ \star_6\Omega =-i\Omega  ~;~~~\star_6\Phi^{(2,1)}=i\Phi^{(2,1)}~.}
\eq{\label{volstuff}\Omega \swed \Omega^*=-\frac{i}{6}|\Omega|^2~\!\text{vol}_6~;~~~
\Phi \swed \Phi^*=\frac{i}{6}|\Phi|^2~\!\text{vol}_6~;~~~
\text{vol}_6=-\frac16 J^3
~,}
where the Hodge star is with respect to the metric of the internal six-dimensional manifold. 
In the case of a CY, $\Omega$, $\Phi$ are harmonic, so their norms are constant (independent 
of the CY coordinates). From \eqref{volstuff} it then follows,  
\eq{\label{volstuff2}\int\Omega \swed \Omega^*=-\frac{i}{6}|\Omega|^2V_6~;~~~
\int\Phi \swed \Phi^*=\frac{i}{6}|\Phi|^2V_6
~,}
where ${V}_6:=\int\text{vol}_6$ is constant, 
cf.~the discussion below \eqref{ans:metric}.

Our spinor conventions  are as in \cite{Terrisse:2019usq}. A Euclidean 4d Weyl spinor of positive, negative chirality  is indicated with a lower, upper spinor index respectively: 
$\theta_\alpha$, $\chi^\alpha$, i.e.~the position  indicates the chirality. 
The 4d gamma matrices, the  charge conjugation and chirality matrices are decomposed into chiral blocks,
\eq{\label{a1}
\gamma_\mu=
\left( {\begin{array}{cc}
   0 & \left(\gamma_\mu\right)_{\alpha\beta} \\
   \left(\gamma_\mu\right)^{\alpha\beta} & 0
  \end{array} } \right)~;~~~
  C^{-1}=
  \left( {\begin{array}{cc}
  C^{\alpha\beta} & 0 \\
  0 & C_{\alpha\beta}
  \end{array} } \right)~;~~~
    \gamma_5=
  \left( {\begin{array}{cc}
  \delta_{\alpha}{}^{\beta} & 0 \\
  0 & -\delta^{\alpha}{}_{\beta}
  \end{array} } \right)
~.}
The $(C^{-1}\gamma_{\mu_1\dots\mu_n})$'s possess the following symmetry properties,
\eq{\label{symprp}
\left(C^{-1}\gamma_\mu\right)_\alpha{}^{\beta}=(-1)^{\frac{1}{2}(n-2)(n-3)}\left(C^{-1}\gamma_\mu\right)^{\beta}{}_{\alpha}
~.}
The Fierz relation for two positive-chirality 4d spinors reads,
\eq{\label{a4}
\theta_\alpha\chi_\beta=-\frac12(\tilde{\theta}\chi)C_{\alpha\beta}-\frac18(\tilde{\theta}\gamma_{\mu\nu}\chi)\left(\gamma^{\mu\nu}C\right)_{\alpha\beta}
~,}
with a similar relation for negative-chirality spinors, and we have defined $\tilde{\theta}:= \theta^{\text{Tr}}C^{-1}$. 
Let us also mention that 
for any two 4d positive-chirality traceless vector-spinors, $\theta_+^\mu$, $\chi_+^\mu$, we have,
\eq{\label{fgt}
\big(\theta^{[\mu_1}_+\gamma^{\mu_2\mu_3}\chi^{\mu_4]}_+\big)=\frac{1}{12} \varepsilon^{ \mu_1\mu_2\mu_3\mu_4 }
\big(\theta^{\lambda}_+\chi_{\rho+}\big)
~.
}

\section{The 10d theory}\label{sec:the10d}

In the conventions of \cite{Terrisse:2018qjm}, upon setting the dilatino to zero, the  IIA action  reads, 
\eq{\spl{\label{action3}
S=S_b&+\frac{1}{2\kappa_{10}^2}
\int\d^{10}x\sqrt{{g}} \Big\{ 
2(\tilde{\Psi}_M\Gamma^{MNP}\nabla_N\Psi_P)
+\frac{1}{2}e^{5\phi/4}m(\tilde{\Psi}_M\Gamma^{MN}\Psi_N) \\
&-\frac{1}{2\cdot 2!}e^{3\phi/4} F_{M_1M_2}(\tilde{\Psi}^M\Gamma_{[M}\Gamma^{M_1 M_2}\Gamma_{N]}\Gamma_{11}\Psi^N) \\
&-\frac{1}{2\cdot 3!}e^{-\phi/2} H_{M_1\dots M_3}(\tilde{\Psi}^M\Gamma_{[M}\Gamma^{M_1\dots M_3}\Gamma_{N]}\Gamma_{11}\Psi^N) \\
&+\frac{1}{2\cdot 4!}e^{\phi/4} G_{M_1\dots M_4}(\tilde{\Psi}^M\Gamma_{[M}\Gamma^{M_1\dots M_4}\Gamma_{N]}\Psi^N) 
+L_{\Psi^4}\Big\}
~,}}
where $\Psi_M$ is the gravitino; $S_b$ denotes the bosonic sector,  
\eq{\spl{\label{ba}S_b= \frac{1}{2\kappa_{10}^2}\int\d^{10}x\sqrt{{g}}\Big(
&-{R}+\frac12 (\partial\phi)^2+\frac{1}{2\cdot 2!}e^{3\phi/2}F^2\\
&+\frac{1}{2\cdot 3!}e^{-\phi}H^2+\frac{1}{2\cdot 4!}e^{\phi/2}G^2
+\frac{1}{2}m^2e^{5\phi/2}\Big) 
+S^\mathrm{CS}
~,
}}
and  $S^\mathrm{CS}$ is the Chern-Simons term, which in the massless limit is simplifies to,
\eq{
S^\mathrm{CS}=\int \big(\tfrac12B\swed G^2-\tfrac12 F \swed B^2\swed  G+\tfrac16 F^2 \swed B^3\big)
~.}
$L_{\Psi^4}$ in \eqref{action3} denotes the 24  quartic gravitino  terms given in \cite{Giani:1984wc}. 
Of these only four can have a nonvanishing VEV in a gravitational instanton background with positive-chirality 4d zero modes, resulting in the expression given in  
\eqref{calbdef}.

We emphasize that the action \eqref{action3}  should 
be regarded as a 
book-keeping device whose variation with respect to the bosonic fields gives the correct bosonic
equations of motion in the presence of gravitino condensates. 
Furthermore, the
fermionic equations of motion are trivially satisfied in the maximally-symmetric vacuum. 
The  (bosonic) equations of motion (EOM) following from  (\ref{action3}) are as follows:

Dilaton EOM,
\eq{\spl{\label{beomf1}
0&=-{\nabla}^2\phi+\frac{3}{8}e^{3\phi/2}F^2-\frac{1}{12}e^{-\phi}H^2+\frac{1}{96}e^{\phi/2}G^2 +\frac{5}{4}m^2e^{5\phi/2}\\
&+\frac{5}{8}e^{5\phi/4}m(\tilde{\Psi}_M\Gamma^{MN}\Psi_N) \\
&-\frac{3}{16}e^{3\phi/4} F_{M_1M_2}(\tilde{\Psi}^M\Gamma_{[M}\Gamma^{M_1 M_2}\Gamma_{N]}\Gamma_{11}\Psi^N)\\
&+\frac{1}{24}e^{-\phi/2} H_{M_1\dots M_3}(\tilde{\Psi}^M\Gamma_{[M}\Gamma^{M_1\dots M_3}\Gamma_{N]}\Gamma_{11}\Psi^N) \\
&+\frac{1}{192}e^{\phi/4} G_{M_1\dots M_4}(\tilde{\Psi}^M\Gamma_{[M}\Gamma^{M_1\dots M_4}\Gamma_{N]}\Psi^N) 
~.
}}
Einstein EOM,
\eq{\spl{\label{beomf2}
{R}_{MN}&=\frac{1}{2}\partial_M\phi\partial_N\phi+\frac{1}{16}m^2e^{5\phi/2}{g}_{MN}
+\frac{1}{4}e^{3\phi/2}\Big(  2F^2_{MN} -\frac{1}{8} {g}_{MN}  F^2 \Big)\\
&+\frac{1}{12}e^{-\phi}\Big(  3H^2_{MN} -\frac{1}{4} {g}_{MN}  H^2 \Big)
+\frac{1}{48}e^{\phi/2}\Big(   4G^2_{MN} -\frac{3}{8} {g}_{MN}  G^2 \Big)\\
&+ \frac{1}{24}e^{\phi/4}G_{(M|}{}^{M_1M_2 M_3} 
 (\tilde{\Psi}_P\Gamma^{[P}\Gamma_{|N)M_1M_2 M_3}\Gamma^{Q]}\Psi_Q) \\
&-\frac{1}{96}e^{\phi/4}G_{M_1\dots M_4}\Big\{
(\tilde{\Psi}_P\Gamma_{(M}\Gamma^{M_1\dots M_4}\Gamma^{P}\Psi_{N)})-(\tilde{\Psi}_P\Gamma^P\Gamma^{M_1\dots M_4}\Gamma_{(M}\Psi_{N)})\\
&+\frac12 g_{MN}(\tilde{\Psi}^P\Gamma_{[P}\Gamma^{M_1\dots M_4}\Gamma_{Q]}\Psi^Q) 
\Big\}
-\frac18 g_{MN}L_{\Psi^4}+\frac{\delta L_{\Psi^4}}{\delta g^{MN}}
~,}}
where we have set: $\Phi^2_{MN}:=\Phi_{MM_2\dots M_p}\Phi_N{}^{M_2\dots M_p}$, for any $p$-form $\Phi$. 
In the Einstein equation above we have not included the gravitino couplings to the two- and three-forms, which vanish in the   background given by \eqref{ans:gravitino}. 

Form EOM's,
\eq{\spl{\label{beomf3}
0&=\d {\star}\big[ e^{3\phi/2}F  
-\frac{1}{2}e^{3\phi/4} (\tilde{\Psi}^M\Gamma_{[M}\Gamma^{(2)}\Gamma_{N]}\Gamma_{11}\Psi^N)
\big]+ H\swed {\star} \big[e^{\phi/2}G {+ \frac{1}{2}e^{\phi/4}  (\tilde{\Psi}^M\Gamma_{[M}\Gamma^{(4)}\Gamma_{N]}\Psi^N)} \big]\\
0 &= \d{\star} \big[ e^{-\phi}H
-\frac{1}{2}e^{-\phi/2}  (\tilde{\Psi}^M\Gamma_{[M}\Gamma^{(3)}\Gamma_{N]}\Gamma_{11}\Psi^N)
\big]
+e^{\phi/2}F\swed {\star} \big[e^{\phi/2}G { + \frac{1}{2}e^{\phi/4}  (\tilde{\Psi}^M\Gamma_{[M}\Gamma^{(4)}\Gamma_{N]}\Psi^N)} \big]\\
 & -\frac{1}{2}G\swed G
+ m {\star}\big[ e^{3\phi/2}F  
{-\frac{1}{2}e^{3\phi/4} (\tilde{\Psi}^M\Gamma_{[M}\Gamma^{(2)}\Gamma_{N]}\Gamma_{11}\Psi^N)}
\big]\\
0&=\d
{\star} 
\big[
e^{\phi/2}G
   +\frac{1}{2}e^{\phi/4}  (\tilde{\Psi}^M\Gamma_{[M}\Gamma^{(4)}\Gamma_{N]}\Psi^N) \big]
-H\swed G
~,
}}
where
$\Gamma^{(p)}:=\frac{1}{p!}\Gamma_{M_1\dots M_p}\d x^{M_p}\wedge\dots\wedge\d x^{M_1}$. In addition the forms obey the  Bianchi identities,
\eq{\label{bi}
\d F= mH~;~~~\d H=0~;~~~\d G=H\wedge F
~.}

\section{Gauged $\mathcal{N}=2$, 4d supergravity}\label{sec:gaugedsugra}

Massless IIA compactification with background fluxes on a CY, $Y$, with Hodge numbers $(h^{1,1},h^{2,1})$, results in gauged $\mathcal{N}=2$ supergravity with 
$n_V=h^{1,1}$ vector multiplets and $n_H=h^{2,1}+1$ hypermultiplets. 
The vector multiplets contain $n_V$ complex scalars $t^i$, $i=1,\dots,n_V$, arising from 
the expansion of the complexified K\"{a}hler form,
\eq{\label{99}B+i J
=t^i \omega_i
~,}
on a basis $\{\omega_i\}$ of real harmonic (1,1)-forms of the CY. Let us introduce the following definitions,
\eq{\label{100}
\kcal=\frac16\int_Y J\swed J\swed J~;~~~
\kcal_i= \int_Y \omega_i\swed J\swed J~;~~~
\kcal_{ij}=\int_Y \omega_i\swed \omega_j\swed J~;~~~
\kcal_{ijk}= \int_Y \omega_i\swed \omega_j\swed \omega_k
~.}
The associated  K\"{a}hler potential is then given by,
\eq{\label{101}
e^{-K_V}=8\kcal~.}
We will also need the symmetric matrix $\ncal_{IJ}$, $I,J=0,1,\dots, n_V$, whose components can be expressed as follows,
\eq{\spl{\label{37}
\Re\ncal_{00}&=-\frac13 \kcal_{ijk}
\Re t^i \Re t^j \Re t^k~;~~~
\Re\ncal_{i0}=\frac12 \kcal_{ijk}
 \Re t^j \Re t^k~;~~~
\Re\ncal_{ij}=- \kcal_{ijk}
  \Re t^k\\
 \Im\ncal_{00}&=-\kcal+\left(\kcal_{ij}-    \frac{\kcal_i\kcal_j}{4\kcal}\right)
\Re t^i \Re t^j ~;~~~
 \Im\ncal_{i0}=-\left(\kcal_{ij}-    \frac{\kcal_i\kcal_j}{4\kcal}\right)
 \Re t^j ~;~~~
 \Im\ncal_{ij}=\kcal_{ij}-  \frac{\kcal_i\kcal_j}{4\kcal}
~.}}
In order to describe the hypermultiplets it is useful to introduce a real basis $\{\alpha_A,\beta^B\}$ of harmonic three-forms on $Y$ normalized as follows,
\eq{\spl{\label{3basis}
\int_Y\alpha_A\swed\beta^B=\delta_A^B~;~~~
\int_Y\alpha_A\swed\alpha_B=\int_Y\beta^A\swed\beta^B=0
~,}}
for $A,B=0,1,\dots, h^{2,1}$ . The periods of $\Omega$ are then given by,
\eq{\label{104}
%Z^A=\int_Y\Omega\swed\beta^A~;~~~
%\gcal_A=\int_Y\Omega\swed\alpha_A~;~~~
\Omega=Z^A\alpha_A-\gcal_A\beta^A
~,}
where $\mathcal{G}_A={\partial\mathcal{G}}/{\partial Z^A}$, with $\mathcal{G}(Z)$ a holomorphic, homogeneous, degree-two function of the projective coordinates $Z^A$. 
The complex structure 
deformations are parameterized by the affine coordinates $z^a=Z^a/Z^0$, with $a=1,\dots,h^{2,1}$. This is equivalent to setting $Z^A=(1,z^a)$. 
The metric on the space of complex structure deformations $g_{a\bar{b}}=\partial_{z^{a\phantom{b}}}\!\!{\partial}_{\bar{z}^b}K_H$ is associated with the K\"{a}hler potential,
\eq{\label{105}
K_H=-\ln \Big(i\int_Y\Omega\swed\bar{\Omega}\Big)
=-\ln i \left(
\bar{Z}^A\mathcal{G}_A-Z^A\bar{\mathcal{G}}_A
\right)
~.}
We will also need to define the symmetric matrix $\mcal$,  
\eq{\label{106}
\mcal_{AB}:=\bar{\gcal}_{AB} +2i\frac{(\Im \gcal)_{AC} Z^C  (\Im \gcal)_{BD}Z^D}{Z^C  (\Im \gcal)_{CD}Z^D}
~,}
where $\mathcal{G}_{AB}:={\partial^2\mathcal{G}}/{\partial Z^A\partial Z^B}$.

The matrix $\mcal_{AB}$ can be used to express the Hodge dual of the three-form basis,
\eq{\label{abstar}
\star_6\alpha_A=A_{A}{}^B\alpha_B+B_{AB}\beta^B~;~~~
\star_6\beta^A=C^{AB}\alpha_B-\beta^BA_{B}{}^A
~,}
where $A_{A}{}^B$, $B_{AB}$,  $C^{AB}$ are given by, 
\eq{\spl{
A=\Re \mcal\cdot \Im \mcal^{-1}~;~~~
B=-\Im \mcal  -\Re \mcal  \cdot\Im \mcal^{-1} \cdot\Re \mcal~;~~~
C=\Im\mcal^{-1}
~,}}
and we have used matrix notation.

We shall be interested in the bosonic sector of $\mathcal{N}=2$ supergravity 
with $n_V$  ungauged vector multiplets and $n_H$ gauged hypermultiplets, 
\eq{\spl{\label{n2}
S_4=\int&\d^4 x\sqrt{g_4}
\Big(
R
  - 2g_{ij}\partial_{\mu} t^i\partial^{\mu} \bar{t}^j-2h_{uv}D_\mu q^uD^\mu q^v
 +\Im{\ncal}_{IJ}F^I_{\mu\nu} F^{J\mu\nu}-V
 \Big)
 +\int \Re{\ncal}_{IJ}F^I\swed F^J
~,
}}
where $u,v=1,\dots, 4n_H$. The metric $h_{uv}$ reads,
\eq{\spl{\label{qmetric}
h_{uv}Dq^uDq^v&=
(\d\hat{\phi})^2+ \tfrac{1}{4} e^{4\hat{\phi}} (D b)^2+g_{a\bar{b}}\d z^a\d \bar{z}^b\\
&- \tfrac{1}{4} e^{2\hat{\phi}} (\Im \mcal^{-1})^{AB}\left(
D{\xi}'_A+\mcal_{AC}\d\xi^C
\right)
\left(
D{\xi}'_B+ {\mcal}^*_{BD}\d\xi^D
\right)
~,}}
where  the covariant derivatives, 
\eq{\label{43prime}
Dq^u=\d q^u-k^u_IA^I
~,} 
with $\d A^I=F^I$, are given in terms of the Killing vectors $k^u_I$. 
The potential of the theory is expressed in terms of the latter as follows,
\eq{\label{potac}
V=8e^KX^I\bar{X}^Jh_{uv}k^u_Ik^v_J-\Big(
8e^KX^I\bar{X}^J+(\Im \ncal^{-1})^{IJ}
\Big)\vec{P}_I\cdot \vec{P}_J
~,}
where the $X^I$ are holomorphic functions of the complexified K\"{a}hler coordinates, and may be chosen as $X^I=(1,t^i)$. The prepotentials $\vec{P}_I$ is given as follows: the metric $h_{uv}$ is quaternionic and possesses an associated $SU(2)$ connection $\vec{\omega}=(\omega^1,\omega^2,\omega^3)$,
\eq{\spl{\label{prep1}
\omega^1&=-\tfrac{1}{\sqrt{2}}e^{\frac{1}{2}K_H+\hat{\phi}}  Z^A
\left( \d{\xi}'_A+\mcal_{AB}\d\xi^B
\right)+\text{c.c.}\\
\omega^2&=\tfrac{1}{\sqrt{2}}ie^{\frac{1}{2}K_H+\hat{\phi}}  Z^A
\left( \d{\xi}'_A+\mcal_{AB}\d\xi^B
\right)+\text{c.c.}\\
\omega^3&=\tfrac{1}{\sqrt{2}} e^{2\hat{\phi}} \Big(\d b+\tfrac12(\xi^A\d\xi'_A-\xi_A'\d\xi^A)\Big)+\dots
~,}}
where the ellipses denote the directions along $\d z^a$. We then have \cite{Michelson:1996pn},
\eq{\label{vecp}\vec{P}_I=k^u_I\vec{\omega}_u~.}

\end{document}